# Role of Wettability, Adhesion, and Instabilities in Transitions During Lubricated Sliding Friction

Hao Dong[1], Reshma Siddiquie[2], Xuemei Xiao[3], Michael Andrews[4], Brian Bergman[4], Chung-Yuen Hui[3], Anand Jagota[1,2,*]

[1]Department of Chemical and Biomolecular Engineering, Lehigh University, Bethlehem, PA 18015, USA
[2]Department of Bioengineering, Lehigh University, Bethlehem, PA 18015, USA
[3]Department of Mechanical and Aerospace Engineering, Cornell University, Ithaca, NY 14853, USA
[4]Michelin Americas Research Center, Michelin North America Inc., Greenville, SC 29605, USA

* Corresponding author: E-mail: anj6@lehigh.edu

## Abstract

Lubricated contacts in soft materials are important in various engineering systems and natural settings. Three major lubrication regimes are boundary (BL), mixed (ML), and elasto-hydrodynamic (EHL) lubrication, where the contact region is dry, partially wetted, or fully wetted, respectively. The transition between these regimes is insufficiently understood, especially for soft contacts, which impedes desired control of lubricated sliding friction. Here, we report on the role of solid wettability and adhesion on these transitions. Wettability of glycerol on polydimethylsiloxane (PDMS) surface, and adhesion between a glass indenter and PDMS, were varied by exposure of the PDMS to an ultraviolet light-ozone (UV-Ozone) cleaner. By combining friction tests and visualization, we demonstrate that the transition from ML to BL regime is dominated by the wettability of the lubricant; increasing wettability of glycerol makes removal of liquid from the contact region more difficult. Transition from EHL to ML is related to a series of events with increasing normal load, which are thinning of the lubricant layer, sudden jump to contact between the glass indenter and solid substrate across a gap of tens to a few hundreds of nanometers, and attendant elastic instabilities such as wrinkling and stick-slip. These results provide a deeper understanding of transitions in lubricated frictional behavior of soft materials which govern the maximum and minimum friction achievable.

## 1. Introduction

Soft lubricated contacts involve one or two compliant solid surfaces with a liquid layer in-between and are important in various engineering and natural settings, such as submarine grippers[1], soft robots[2-4], tires on wet roads[5-7], haptic applications[8], flexible electronic[9] and energy harvesting devices[10,11]. Although the control of lubricated sliding friction is highly desired, the complicated interplay between flow of the liquid lubricant and deformation of the two solid surfaces makes the manipulation of friction a challenge.

Numerous factors such as normal loading[12-14], substrate stiffness[14,15], viscoelasticity[16], surface patterning[12,17-19], sliding velocity[13,19], and lubricant viscosity[20,21] can influence the frictional response of a system. It has long been established that lubricated frictional response tends to lie in one of three regimes.[22] These regimes are Boundary (BL), Mixed (ML), and Elasto-hydrodynamic (EHL) lubrication in which the contact between two solid surfaces is dry, partially wetted, or fully wetted, respectively[22]. Within a single lubrication regime friction usually changes monotonically with, for example, sliding velocity, facilitating prediction of friction behavior within that regime. Transition between lubrication regimes is usually accompanied by a change of the monotonic trend. Consequently, these transitions correspond to extreme friction values achievable in a given system[21] and are thus very important. For example, the transition from EHL to ML usually corresponds to minimal friction while the transition from ML to BL usually corresponds to maximal friction.

Whether the transition from ML to BL is achievable depends on whether the lubricant can be fully squeezed out[23]. For example, a lubricant with poor wettability on a solid substrate (e.g. water on polydimethylsiloxane (PDMS)) can be fully removed from the contact region, while it may not be possible to expel a lubricant with good wettability to the solid substrate (e.g. silicone oil on PDMS)[21,24,25].

Numerous endeavors have been made to understand how the EHL-ML transition[21, 26] is influenced by surface roughness[14, 27], surface structure[17], stiffness of the substrates[21], viscosity or type of the lubricant[28] and lubricant film thickness[21]. The generally accepted criterion for the transition is that it occurs when some measure of surface roughness amplitude is on the order of the lubricant film thickness[15]. However, a recent study showed that the EHL to ML transition in smooth soft materials usually occurs for film thickness much larger than the roughness amplitude[21]. Little is understood about how solid-solid contact is initiated, nor about the origin and influence of various attendant mechanical instabilities.

As a result of low stiffness, compliant materials often suffer large deformation, and their mechanical behavior is consequently often non-linear. This tends to lead to surface elastic instabilities when compliant materials are under sufficiently large deformation[29]. A wide range of instability morphologies such as folding[29], buckling[30], creasing[31], stick-slip[32] and Schallamach waves[33] have been reported. A recent study suggests that there is a long-range (~100 nm) attractive force between two initially separated surfaces, which can cause sudden jump to solid-solid contact even with the existence of an intervening water layer[34].

In this work, we study the effect of lubricant wettability and solid-solid adhesion on the ML-BL and EHL-ML transitions. We use an ultraviolet light-ozone (UV-Ozone) cleaner to treat a PDMS surface to enhance its wettability by glycerol and simultaneously to change the adhesion between the glass indenter and PDMS surface. In lubricated sliding friction tests, we demonstrate that increasing wettability of glycerol on PDMS substrates shifts ML-BL transitions to higher normal load. Indentation tests accompanied by visualization show directly that higher wettability makes it more difficult to squeeze glycerol out of the contact region. We propose that the EHL-ML transition is related to a series of events that occur in sequential order with increasing normal load (1) the lubricant film gets thinner; (2) a portion of the soft solid jumps across the fluid gap to initiate solid-solid contact with the substrate, which causes an increase in friction; (3) other surface elastic instabilities such as wrinkling and stick-slip arise. Our results reveal how wettability and adhesion contribute to the lubricated sliding friction transitions and provide new understanding of lubricated sliding processes for soft materials.

## 2. Results and Discussion

### 2.1 Wettability of lubricant and adhesion between solid surfaces

To investigate the wettability of glycerol on PDMS substrates, the steady-state advancing contact angle (see Supplementary Note 1) of glycerol was measured by an optical goniometer[35,36]. Figures 1(a,b) show the contact angle measurement results for glycerol on stiff (shear modulus ~ 1 MPa) and soft PDMS (shear modulus ~ 40 kPa) substrates[12] treated by a UV-Ozone cleaner for varying time duration. It is clear that for both stiff and soft PDMS substrates, increasing UV-Ozone treatment time yields smaller contact angle. The contact angle distribution on soft PDMS is broader, possibly due to its lower degree of cross-linking[37]. Moreover, longer treatment increases scatter, possibly due to the treatment introducing some degree of surface heterogeneity.

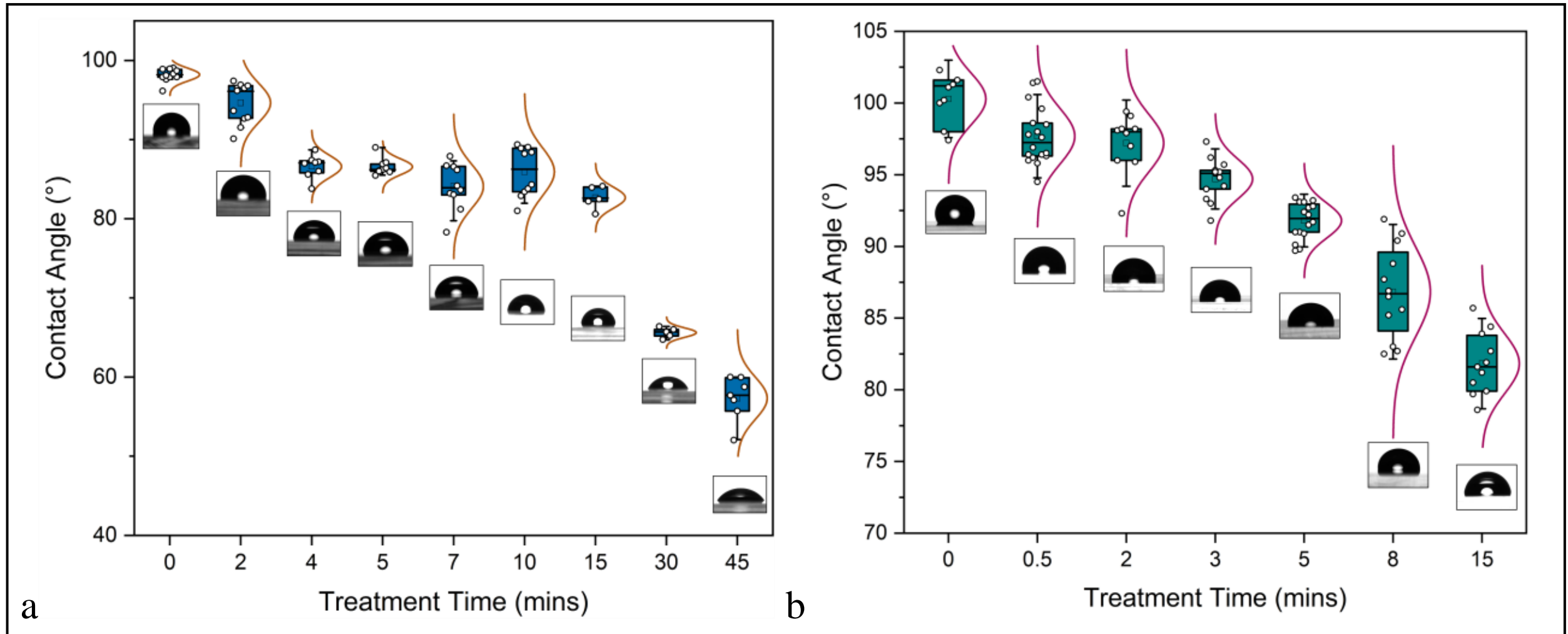

**Figure 1| Steady state contact angle measurement of glycerol on PDMS samples. a**,**b**, statistics of glycerol contact angle on stiff (**a**) and soft (**b**) PDMS substrates. The curve next to each box plot in (**a**) and (**b**) shows the normal distribution of contact angle data. The inserted pictures show examples of the glycerol droplets.

We also characterized the dry and wet (by glycerol) adhesion between the glass indenter and stiff or soft treated PDMS substrates by measuring pull-off force via an indentation setup, Figure 2 (See Experimental Section for details). Evidently, UV-Ozone treatment initially significantly increases the adhesion energy under dry conditions. This increase eventually peaks and decreases with further increase in treatment time[38]. The reduction of dry work of adhesion after 5 mins treatment is probably caused by the formation of a glassy layer on the surface[39,40]. The wet adhesion work shows a similar trend except that its maximum value is much smaller and starts to decrease at shorter treatment time. That the wet adhesion energy becomes vanishingly small when the treatment time is sufficiently long (> 8mins) implies that glycerol on those soft PDMS substrates is very difficult to remove completely. To quantitatively characterize the difficulty of removing glycerol on treated PDMS surfaces, we introduce a surface coverage parameter $\phi$ to represent the proportion of contact area that is wetted by the liquid during a loading and unloading cycle (derivations are in Supplementary Note 2). The parameter $\phi$ is, assuming the wetted portion has much lower adhesion,

$$\phi = (1 - \frac{W_{a,wet}}{\Delta A}) \quad (1)$$

where $W_{a,wet}$ is the adhesion energy measured under wet conditions and $\Delta A$ is the thermodynamic energy required to open a solid-solid interface and fully wet the two opened solid surfaces. As shown in Fig. 2a, $\phi$ increases with increasing treatment time, which corresponds to more liquid on the solid surface.

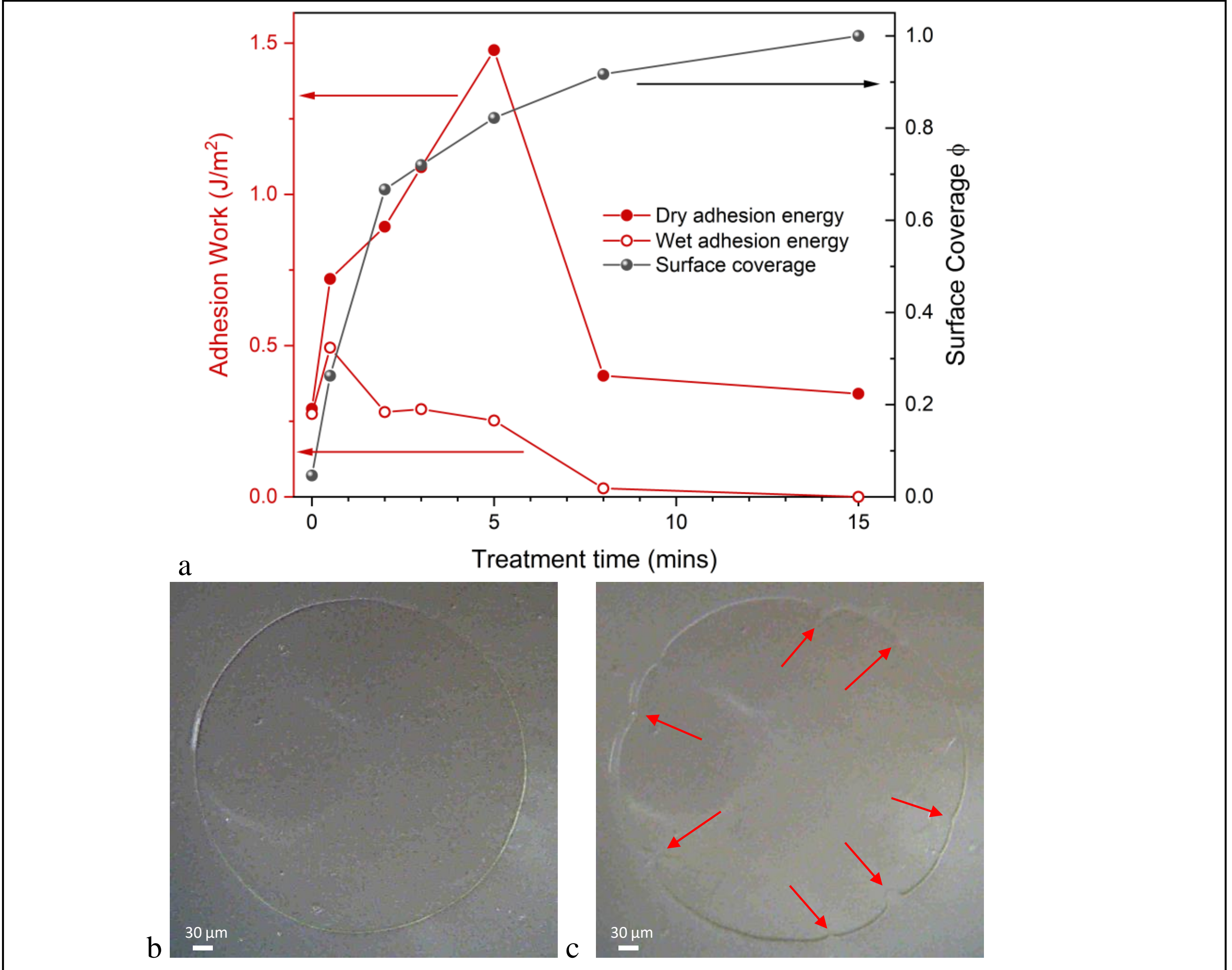

**Figure 2| Adhesion energy and images of contact region. a**, Adhesion energy of a glass indenter on soft PDMS substrates and effective surface coverage $\phi$ versus UV-Ozone treatment time. The radius of the glass indenter and separation velocity are 2.03 mm and 1 μm/s, respectively. The indentation depth for each case is approximately 0.3 mm. **b,c,** Optical micrographs of the contact region between a glass indenter and soft PDMS substrates treated by UV-Ozone for (**b**) 0 min and (**c**) 5 mins. The images were captured during the separation process. Red arrows in (**c**) indicate liquid pockets present in the contact region. The contact edge in (**b**) is sharp, while the contact edge in (**c**) shows the traces of glycerol residue, implying that glycerol becomes more difficult to remove from the contact region with increasing UV-Ozone treatment time. See also Fig. S2 in Supplementary Information.

To demonstrate that glycerol is difficult to remove from PDMS surfaces with enhanced wettability, we visualized the separation process between the glass indenter and PDMS substrates with glycerol as lubricant. Figure 2b, c and Supplementary Figure S2 show images during the separation process (unloading) of a glass indenter with the PDMS substrates treated by UV-Ozone for 0, 5 and 15 mins, respectively. The contact region is a perfect circle for untreated PDMS. However, for 5-mins treatment, the contact edge is not a circle anymore and we observe some liquid entrainment near the contact edge (indicated by the red arrows in Fig. 2c). Furthermore, the contact edge essentially disappears if the surface is treated for 15 mins (Fig. S2d) because the solid surfaces remain separated by a layer of glycerol. This provides direct evidence that glycerol cannot be easily removed from PDMS surfaces with strong wettability. As a result, the lubrication regime transitions (from ML to BL or from EHL to ML regimes) occur at larger normal load. Also, the BL regime might not be achievable for surfaces with high wettability.

**2.2 Lubricated sliding friction of PDMS treated by UV-Ozone**

To demonstrate the impact of glycerol wettability on lubricated sliding friction, we plot normalized force vs normalized velocity using a normalization from isoviscous EHL theory[13]. EHL theory shows that in the EHL regime the normalized friction F:

$$F = fR^{2/3}G^{1/3}N^{-4/3} \,, \tag{2}$$

depends on a single dimensionless parameter V (normalized velocity) given by

$$V = v\eta R^{5/3}G^{1/3}N^{-4/3}, \tag{3}$$

where *f* is the friction force, $\eta$ is the lubricant viscosity, v is the sliding velocity, *R* is the radius of the indenter, *G* is the shear modulus of the substrate, and *N* is the normal load.

Figures 3a and b plot *F* versus *V* for low sliding velocity cases (0.1 mm/s) on stiff and soft PDMS. *F* versus *V* for higher velocity cases (0.5 and 1 mm/s) are shown in Figs. S4. From Fig. 3 it is evident that the UV-Ozone treatment has a major influence on the sliding friction measurements. For stiff PDMS substrates, as the wettability of glycerol is enhanced (UV-Ozone treatment time increases), the system transitions from BL to ML or from ML to EHL under larger normal load (smaller $V$, Fig. 3a). Moreover, longer treating time (> 7mins) makes glycerol behave like silicone oil of similar viscosity, which is a highly wetting lubricant on untreated PDMS surfaces and can significantly decrease F in the high load region ($V < 10^{-6}$), because the lubricant becomes more difficult to remove.

For soft PDMS systems, the friction behavior of the system (Fig. 3b) becomes more complex. Generally, increasing UV-Ozone treatment time moves the system from BL to ML or ML to EHL regimes for larger normal loads. Sufficiently long treatment time makes glycerol on UV-Ozone treated PDMS similar to silicone oil on untreated PDMS. That the data for dry conditions overlap with those for untreated substrates under BL (smaller V) conditions shows that glycerol can be squeezed out of the contact in BL (Soft PDMS glycerol 0 mins and Soft PDMS dry 0 mins in Fig. 3b). If treated for > 2mins (Fig. 3b), there is now a significant difference between dry and wet friction; glycerol cannot be fully removed.

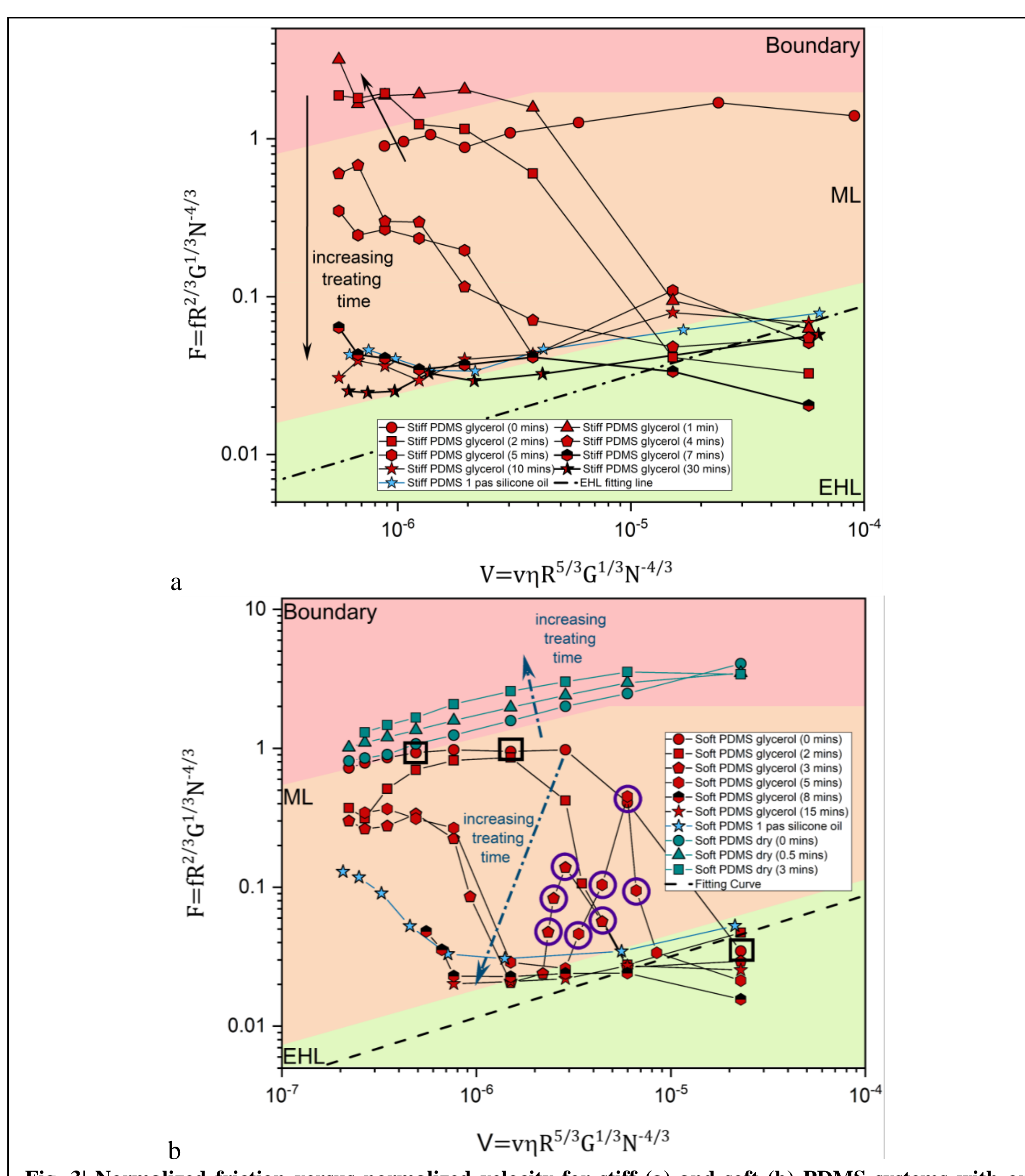

**Fig. 3| Normalized friction versus normalized velocity for stiff (a) and soft (b) PDMS systems with or without UV-Ozone treatment.** The pink, pale-orange, and light green background colors represent BL, ML, and EHL regimes, respectively. The black arrows serve as a guideline to indicate the direction of increasing treatment time. The lubricated sliding velocity, and the indenter radius for all the cases shown here are 0.1 mm/s and 2.03 mm. In the legend, the time indicates duration of UV-Ozone treatment. The data points showing non-monotonic transition behavior are circled in purple. For stiff PDMS substrates, the frictional behavior is mainly controlled by the wettability of the lubricant. With enhanced wettability (increased UV-Ozone treatment time), the normalized friction reduces systematically. For soft PDMS, wettability governs the ML-BL transition. Increasing wettability generally pushes the EHL-ML transition to higher normal load. The non-monotonic transitions in (**b**) suggest a delicate interplay between wettability and adhesion. Sufficiently long treatment (> 7mins in **a** and > 8mins in **b**) makes the glycerol system behave like silicone oil system.

Elaborating further, 2-min treatment shows noticeable friction reduction under large normal load. The lubrication regime transition from ML to BL or EHL to ML regime requires larger loads. Friction in soft samples treated for medium duration of 3-5 mins is reduced for large normal loads due to enhanced wettability. Systems treated for 3-5 mins do not reach BL; dry contact cannot be achieved under the conditions studied here. Interestingly, 3 or 5-min treatment for soft samples results in a double transition, i.e., with increasing N the system goes through lubrication transition from EHL to ML regime twice. As a result, the change in friction with increasing N is strongly non-monotonic. To the best of our knowledge, this is the first report of such a double transition phenomenon. It is closely related to the formation and the morphology of surface elastic instabilities as discussed later. Long-time treatment (> 8 mins), makes glycerol behave like silicone oil on PDMS. As a result, long-term treated samples remain in the EHL regime for most cases and ML regime under high load (≥ 334 mN, Supplementary Video 1).

### 2.3 Surface Instabilities on PDMS Surfaces

A recent study showed that the EHL-ML transition (in untreated PDMS) is accompanied by elastic instabilities[21]. Change in wettability also has an impact on surface instabilities in the contact region. Under dry conditions, as reported previously[41], we observe wavy wrinkles, Schallamach waves, at the leading edge (Fig. 4a, Supplementary Video 2). These wrinkles become air pockets as they are entrained into the contact region[42]. For glycerol-lubricated untreated soft PDMS surfaces, smaller-wavelength wrinkles and glycerol droplets (Fig. 4b, ML, Supplementary Video 3) are observed. These wrinkles trap glycerol and these droplets are dragged into and through the contact region, as previously reported[21]. In the EHL or BL regime, fewer or no glycerol droplets are found. For 2-min UV-Ozone treated surface, the number of the wave-like wrinkles increase dramatically, and many more glycerol droplets arise in the contact region (Fig. S5a, Supplementary Video 4). When the surface is treated for 3-5 mins, the mode of sliding is no longer smooth (Fig. 4c, Supplementary Video 5). Instead, we observe a stick-slip mode of sliding. As shown in Fig. 4c, during the stick stage, a few wrinkles form in the interior of the contact region, not just at the leading edge. These wrinkles adhere strongly enough not to slide with the PDMS sample. Additional shear displacement accumulates in these wrinkles and the friction force increases rapidly. Once the friction force is large enough, these wrinkles go through a sudden release process, the surface flattens, and the force drops to a lower value. The cycle repeats. For long-time (e.g. 15 mins) treated PDMS substrates, we do not find any sign of elastic instabilities (Supplementary Notes 5) since the system is in the EHL regime. For samples treated by UV-Ozone cleaner over 8 mins, sample failure can happen under high normal loads (> 334 mN, see Supplementary Note 5 for more details.)

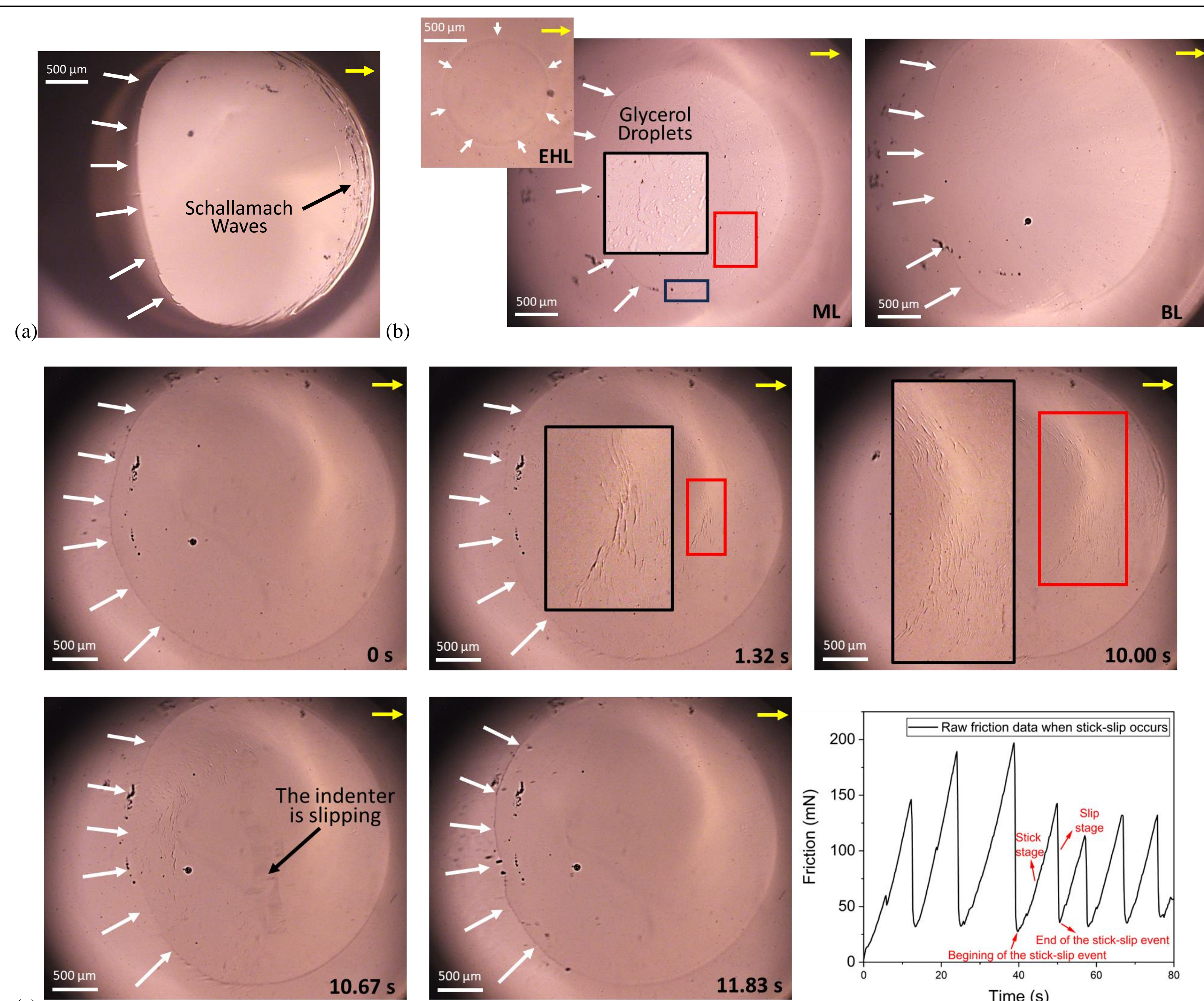

**Figure 4** (a) A still image of an untreated soft PDMS substrate surface under dry condition during sliding. The normal load is 334 mN. (b) Instabilities during lubricated sliding at different normal loads of untreated PDMS. Left: The normal load is 18 mN, and the system is in EHL regime. Sliding is smooth with low friction. Middle: at N = 144 mN, the system is in ML. We observe wrinkles at the leading edge that trap glycerol droplets which get transported across the contact region. Sliding is smooth as is the corresponding friction force. Right: N = 334 mN, the system is in BL. The interface appears to be dry with none to few glycerol droplets. Sliding and friction force are smooth. See black-boxed points in Fig 3b for the corresponding lubrication regime of the three images. (c) Stick-slip instability in softer PDMS samples treated for 3 minutes. The normal load is 334 mN. The time at the lower right corner of each image in (c) indicates the time passed from the beginning of the stick-slip event (first image). In PDMS samples treated by 3 mins, wrinkles form in the interior of the contact region. Portions of the wrinkles adhere strongly to the glass indenter, causing further growth of wrinkles towards the leading edge (1.32-10 s). The shear force rises steeply with time. At a critical condition, the interface slips suddenly, releasing the accumulated shear displacement, the force falls to a lower value abruptly, and the cycle continues. Yellow arrows and white arrows indicate the sliding direction and the edge of the contact region. The black boxed area is the magnified picture of red boxed region. The dark blue boxed area shows a few minor wave-like undulations. Time shown at the right down corner of each image in (c) indicates the time passed after the beginning of the stick-slip event. The videos of the microscopic images in (**a**), (**b,** middle) and (**c**) are Supplementary Videos 2, 3, and 5.

Enhanced wettability tends to keep solid surfaces apart whereas solid-solid adhesion tends to close the interface. The interplay of these two opposing tendencies can result in novel frictional behavior and surface instabilities. To quantify this competition, we propose a parameter ξ

$$\xi = \frac{\gamma_l cos\theta_1 + \gamma_l cos\theta_2}{W_a} \quad , \tag{4}$$

where $W_a$ is the work of solid-solid adhesion of the interface between 1 and 2 in air; $\gamma_l$ is liquid vapor surface energy; $\theta_1$ and $\theta_2$ are contact angles of a liquid on surface 1 and 2. The parameter $\xi$ is the ratio of the spreading ability of the liquid and $W_a$. See SI.6 for a derivation. Figure S6 plots measured $\xi$ versus treatment time. Note the non-monotonicity -- $\xi$ first decreases and then increases with treatment time. For untreated PDMS substrates, the wettability of glycerol on the substrate and solid-solid adhesion are both low. It is easy for glycerol to be expelled from the contact region and a new dry interface to be formed under large normal load and therefore frictional behavior of untreated PDMS at high normal loads is similar to dry PDMS (Fig. 3b). Due to weak adhesion ($\xi$ value is large, Type I in Fig. S6), only a few minor undulations at the leading edge in the contact region are observed (dark blue box in Fig. 4b middle, Supplementary Video 3). For PDMS substrates treated for a short duration (2 mins), wettability is enhanced and dry contact becomes harder to make, and the EHL-ML transition occurs under larger normal load. However, adhesion also increases, and $\xi$ decreases. Solid-solid contact sustains larger loads and the number of wave-like undulations which can carry glycerol droplets into the contact region significantly increases (Supplementary Video 4) and the ML-BL transition occurs under still larger load.

For longer time (3-5 mins) treated systems, ξ reaches its minimum value (Type II, Fig. S6); adhesion grows faster than wettability. In our system, the consequence appears to be that the sliding mode changes to stick-slip (Supplementary Video 5). Fig. 5 plots the non-monotonic change of normalized friction versus normalized velocity for 3-min case with corresponding micrographs of the contact region (see Fig. S7 for 5-min case). The data lie in four regions. Examples of raw data for each region are shown in Fig. S8. In Region I (Supplementary Video 6 and 10), we observe intense stick slip. In Region II (Supplementary Video 7 and 11), the stick-slip vanish, reappearing in Region III (Supplementary Video 8 and 12). Finally, when N is small, Region IV (Supplementary Video 9 and 13), the lubricant film is thick so that adhesion effects are muted, with low friction in the EHL regime. The nonmonotonic friction behavior for 3-5 mins treated systems is closely related to how the solid-solid contact is initiated, which is discussed later.

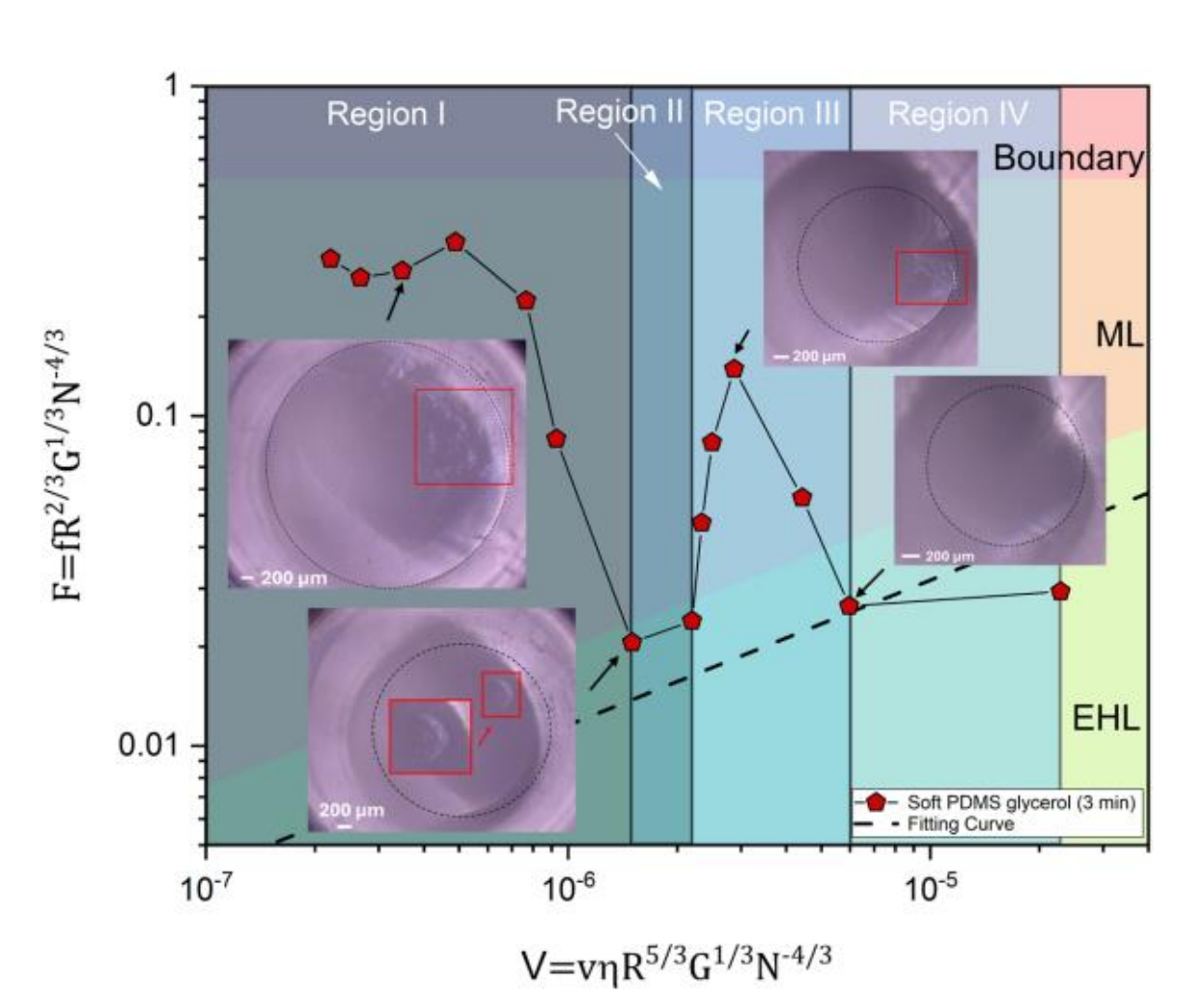

**Fig. 5.** Plot of normalized friction versus normalized velocity for system treated by UV-Ozone for 3 minutes with micrographs of the contact region during sliding. Experimental details: sliding velocity, radius of the indenter, total sliding time (for each direction) and lubricant are 0.1 mm/s, 2.03 mm, 100 s, and glycerol, respectively. The lubricant, lubricated sliding velocity, and the indenter radius for all the cases shown here are glycerol, 0.1 mm/s and 2.03 mm. Inserts: micrographs of the contact region. The black dashed line indicates the contact edge, and the red box highlights stick-slip phenomena in the contact region. The normal load of the inserted images (from left to right) is 334 mN, 144 mN, 88 mN and 51 mN. See supplementary Videos 6-9.

With sufficiently long-time treatment (≥ 8mins), glycerol gets very hard to remove from the contact region, and the system remains in the EHL regime throughout.

### 2.4 Initiation of solid-solid contact

Our results have shown that the lubrication regime transition from EHL to ML is often accompanied by surface instabilities. However, the chronological order of solid-solid contact and surface instabilities remains unknown. A recent study suggests that a long-range attractive force between glass and PDMS can produce possible solid-solid contact even if the two surfaces are separated initially by a thin liquid film[34]. This raises the question: how is solid-solid contact initiated during sliding as one increases normal load? We hypothesize that the critical event is an unstable jump to solid-solid contact[43] when the intervening liquid film gets thin enough, and this event subsequently triggers other instabilities[34].

To support our hypothesis, we utilized a high-speed camera to record the moment when dry contact between a glass indenter and a substrate is just made. When the indenter and the substrate are close enough, the interference patterns (Newton's rings) allow us to calculate the distance between the indenter and the substrate. By fitting the gap size between the indenter and substrate to a parabola, we obtain the distance at which the PDMS suddenly jumps into contact with the indenter (See Experimental Session and Supplementary Notes 9 for details). Figs 6a-b show examples of fitting results with microscope images for stiff and soft PDMS (see Supplementary Notes 10 for all cases). The average sudden jump distances of a glass indenter for stiff and soft PDMS substrates are ($189 \pm 16$) nm and ($224 \pm 13$) nm. Analysis of the image right after the contact, shows that contact was made above the substrate surface (Supplementary Notes 9) in accord with our theoretical model[43].

To investigate the impact of surface chemistry on the jump distance, we performed the same experiments on soft PDMS substrates treated by UV-Ozone cleaner for 5 (Fig. S11e-f) and 15 mins (Fig. 6c). The jump distances for 5- and 15-min cases are ($257 \pm 29$) and ($191 \pm 14$) nm, respectively. The trend of decreasing jump distance with increasing UV-Ozone treatment time is in accordance with our adhesion work measurement (Fig. 2a), which shows that the attractive interaction between the glass indenter and PDMS substrate first increases and then decreases. The same measurement with lubricant in the gap between solid surfaces is more difficult to image. We added DI water as lubricant and found that the jump distance for untreated soft PDMS substrate was reduced to (48±39.1) nm (Fig. 6d and Fig. S11). Evidently the attractive forces are significantly attenuated when a liquid lubricant fills the gap [44]. These experiments support our hypothesis that jumping to contact can occur across gaps significantly greater than the amplitude of surface roughness (2.6 - 4.7 nm, see Experimental Section).

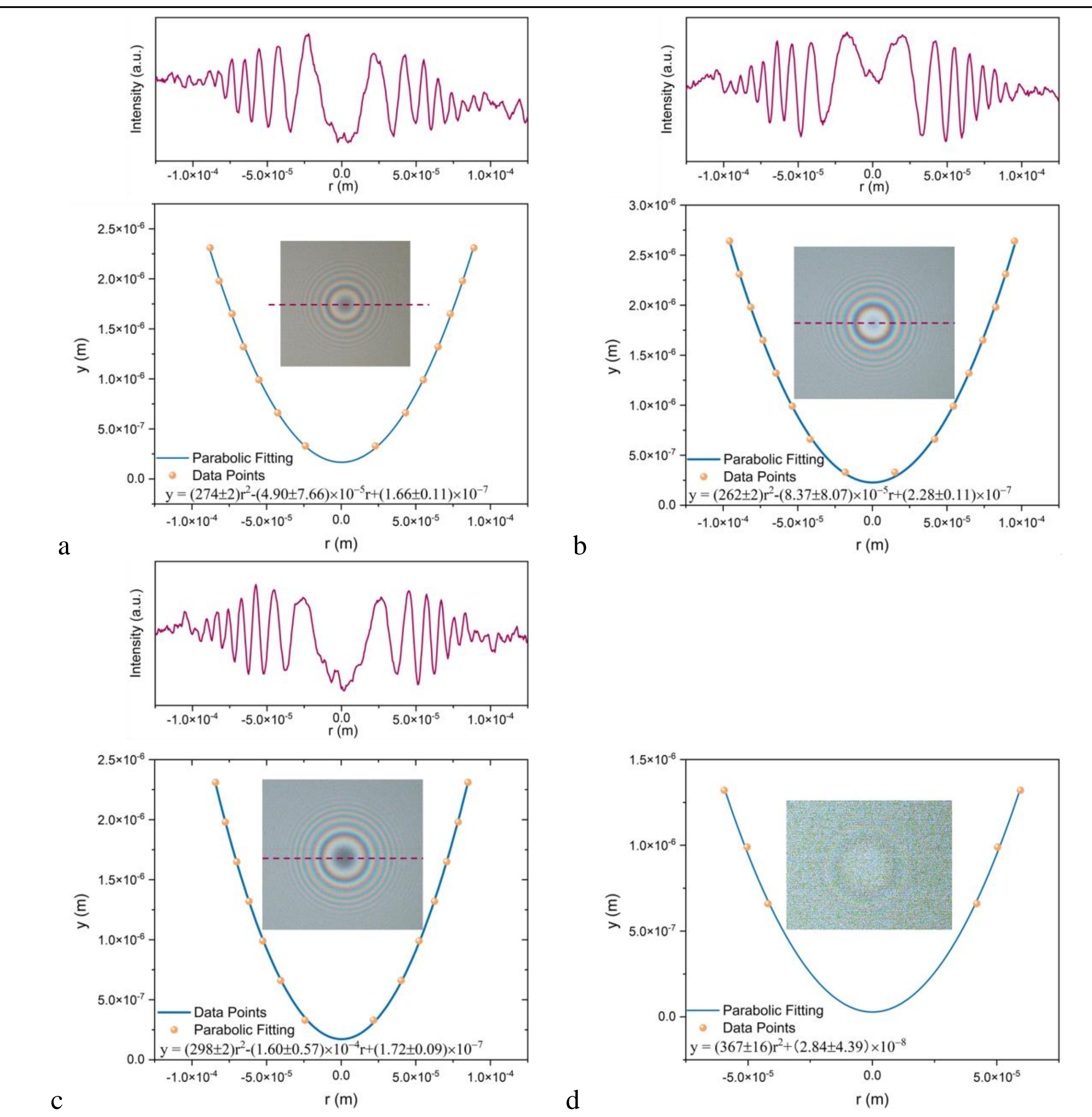

**Fig. 6 Interferometric measurement of the gap between the glass indenter and PDMS substrates just before jump to contact. a,** bottom: gap y between the glass indenter and stiff PDMS (G = 1 MPa) untreated substrates versus horizontal position r just before the contact. Top: the intensity distribution of red light along the horizontal direction indicated by the violet dashed line in the inserted micrograph. **b,c,** bottom: Gap between the glass indenter and soft PDMS (G = 40 kPa) substrates treated by UV-Ozone for (**b**) 0 and (**c**) 15 mins just before the contact. Top: the intensity distribution of red light along the horizontal direction indicated by the violet dashed line in the inserted microscopic images. **d**, gap distance y between the glass indenter and stiff untreated PDMS (G = 1 MPa) just before the contact. In this case, the gap is filled with DI water. The inserts in (**a**), (**b**), (**c**) and (**d**): micrographs of the PDMS surface just before the contact. The dashed lines in (**a**), (**b**), and (**c**) indicate where the cross-section red-light intensity distribution was obtained. It should be noted that because adding water significantly reduced the contrast of the microscopic image, we used a different way to locate the Newton's rings for the water case. See experimental section for details. The parabolic fitting equations can also be found in the figure and the minimum value of y is the jump distance.

Based on the results mentioned above, the observation of two lubrication regime transitions can now be described (Fig. 5). For a system in the EHL regime, the EHL-ML transition with increasing normal load (decreasing *V*) starts with the sudden jump of PDMS into contact with the indenter. For untreated PDMS, we get solid-solid contact but the two sides continue to slide relative to each other smoothly. For treated PDMS the adhesion is large enough to make the solid-solid contact sticky, triggering the stick-slip mode of sliding. With further surface treatment the wettability is too large for the liquid to be removed, resulting in the system returning to a low friction value in the EHL regime.

Thus, the jump into contact across a lubricant-filled gap is the critical event that triggers transition from EHL to ML. We have recently shown[43] that the jump into contact for an elastic foundation with stiffness $c$ (N/m$^3$) separated from a rigid plane by a distance $h$ is controlled by a dimensionless number $\alpha$[43] defined by

$$\alpha = k_f(h)/c \quad (5)$$

where $k_f$ specifies the slope (or stiffness, units in N/m$^3$) of the interaction force (per unit area) vs $h$ defined so that it is positive for attractive interactions. When $\alpha > 1$, a mechanically unstable jump into contact can occur[43]. Equation (5) cannot be used quantitatively for our experiments because it is for normal displacement only. Nevertheless, it explains qualitatively our observations. Consider that the interaction force has two components, a long-range attraction and a short-range repulsion. The attraction could be due to van der Waals forces; the repulsion due to short-range hydration forces. Begin with an untreated glycerol-phobic PDMS surface. The interaction force is purely attractive, and its stiffness increases with decreasing distance such that it eventually exceeds the stiffness of the substrate, then jumping into contact. Surface treatment strengthens the repulsion and the attraction. The effect is that larger normal forces are required to reduce the gap to its critical value. Eventually, the hydration repulsion is large enough to overcome the attraction, so the thickness never reaches the unstable condition.

## 4. Conclusion

We investigated two important regime transitions during lubricated sliding of soft solids: from ML to BL and from EHL to ML regimes, representing the maximum and minimum friction achievable for a given system. We demonstrate that for smooth soft solids, these two transitions depend on the wettability of the lubricant. If the wettability is large enough, the system will stay in the EHL regime and thus the two transition points will not occur even under large normal load. The ML-BL transition, if it exists, is mainly controlled by the wettability of the lubricant. Liquid with high wettability is more difficult to squeeze out from the contact region and hence the ML-BL transition requires a much larger normal load to happen. Our results suggested that the EHL-ML transition follows a typical sequence of events. Holding all other parameters and conditions fixed, imagine starting with low enough normal load for the system to be in the EHL regime. With increasing $N$, at some value the solid surfaces jump into contact. If the surface was treated to increase its adhesion, then that can be large enough to pin the solids to each other. Subsequent sliding is accommodated by production of surface wrinkles each of which is also stuck to the substrate. The force builds up until there is a slip event, a crack-growth kind of instability. If adhesion is insufficiently large, jump into solid-solid contact is followed by steady sliding and the system eventually ends up in BL with increasing $N$. Our results establish several novel features of the phenomenology of lubricated soft solid friction and, as such, will be of broad interest.

## 5. Experimental Section

**Sample Fabrication:** The fabrication methods for PDMS samples have been reported in our previous work[21]. Briefly, PDMS samples were fabricated using a silicone elastomer kit (Dow Sylgard 184, Dow Corning). For the stiff sample, a mass ratio of 10:1 between the base and the curing agent was applied, while for the soft control sample, a mass ratio of 30:1 was used. The elastomer base and curing agent were mixed using a Thicky Mixer and then were cured at 80 °C for 2 h in the oven. The PDMS surfaces as prepared are smooth and have nano scale roughness (Supplementary Notes 11).

**Surface Treatment:** The as-prepared stiff and soft PDMS samples were cut into stripes (5 cm × 1 cm × 2 mm, L × W × H) and the fresh stripes were transferred into a UV-Ozone cleaner (Jelight Model 30) chamber and treated for a certain duration of time (1- 45 mins for stiff PDMS samples and 0.5-15 mins for soft PDMS samples).

**Contact Angle Measurement:** The contact angle of glycerol on PDMS substrates was measured by a goniometer (Rame-Hart Model 210). To get the stable contact angle, the contact angle was measured 40 mins after the glycerol droplets were applied on PDMS surface. Specifically, for soft PDMS treated for 8 and 15 mins, the contact angle data were obtained 4 hours after UV-Ozone cleaner treatment.

**Lubricated Sliding Experiments:** All sliding experiments were done at 24 °C, controlled by the laboratory air-conditioner. Silicone oil (viscosity of 1.0 Pa·s at 24 °C, purchased from Sigma-Adrich), and glycerol (0.99 Pa·s at 24 °C, purchased from Sigma-Adrich) were utilized as lubricants for sliding friction experiments. Lubricant was applied directly on the surface of the PDMS substrates as prepared. The lubricant layer was approximately 1 mm thick. A glass sphere (radius R = 2.03 mm, the fabrication process is discussed later) was brought into contact with the PDMS substrates. The glass sphere was then slid across the lubricated surface under fixed normal load ranging from 19 mN to 605 mN with sliding velocities (controlled by a motor, Newport ESP MFA-CC) of 0.1, 0.5 or 1 mm/s. The sliding friction force was measured by a load cell (Honeywell Precision Miniature Load Cell). The total sliding displacement was 8-12 mm. Lubricated sliding experiments for treated samples were performed 40 mins after UV-Ozone treatment to align with the contact angle measurement. For soft PDMS treated by UV-Ozone cleaner for 8 and 15 mins, the experiments were conducted 4 hours after treatment.

**Glass Indenter Fabrication:** A borosilicate glass rod (diameter = 3 mm) was purchased from McMaster-Carr. The as-purchased glass rod was sintered in the flame of a propane torch, generating an indenter with a smooth spherical tip.

**Adhesion Experiments:** Adhesion experiments were performed on a custom-built indentation setup. For dry adhesion test, a glass sphere (radius R = 2.03 mm) was first brought down until it was roughly 0.1 mm above the PDMS surface. After that, the indenter was brought to contact and continued to indent the PDMS substrates with a vertical velocity of 1 μm/s. The vertical force signal and the position of the indenter were recorded. For wet adhesion tests, the lubricant was first applied on the PDMS surface. Before indentation, the indenter was stopped around 0.1 mm above the PDMS surface and was immersed in the lubricant layer for 30 mins. The remaining steps were the same as for the dry adhesion test. The adhesion energy $W_{ad}$ was calculated based on pull-off force[45]:

$$W_{ad} = \frac{2F_{pull-off}}{3\pi R} \tag{6}$$

where $F_{pull-off}$ is the pull-off force and $R$ is the indenter radius.

**Surface characterization:** The top surface of a PDMS sample as prepared was characterized by an optical profilometer (ZeGage, Zygo). A square region (167 μm × 167 μm) was randomly selected for roughness characterization.

**Jump into Contact Experiment:** see Supplementary Notes 8 for details.

## Acknowledgement

We acknowledge support from the National Science Foundation through the Grant LEAP-HI: CMMI-1854572. The authors thank Prof. Manoj K. Chaudhury for inspiring discussions.